\documentclass[aps,pra,showpacs,twocolumn,a4paper]{revtex4}
\usepackage{amsmath,amsfonts}
\usepackage{graphicx, subfigure}

\newcommand{\id}{\ensuremath{\boldsymbol{1}}}
\newcommand{\ham}{\ensuremath{H}} 
\newcommand{\hilb}{\ensuremath{\mathcal{H}}}

\newcommand{\ke}{\ensuremath{\ket{\text{e}}}}
\newcommand{\be}{\ensuremath{\bra{\text{e}}}}
\newcommand{\kl}{\ensuremath{\ket{1}}}
\newcommand{\ko}{\ensuremath{\ket{0}}}
\newcommand{\kaux}{\ensuremath{\ket{\text{aux}}}}

\DeclareMathOperator*{\im}{Im}
\newcommand{\dd}[1]{\frac{\partial}{\partial{#1}}}
\newcommand{\topp}[1]{^{({#1})}}
\newcommand{\bra}[1]{\left\langle{#1}\right\rvert}
\newcommand{\ket}[1]{\left\lvert{#1}\right\rangle}
\newcommand{\expval}[1]{\left\langle{#1}\right\rangle}

\newcommand{\braket}[1]{\expval{#1}}

\newcommand{\nbar}{\bar{n}}
\newcommand{\pextot}{P\topp{\text{tot}}_e}

\begin{document}

\title{%
Scalable designs for quantum computing with rare-earth-ion-doped crystals}

\date{\today}

\author{Janus H. Wesenberg}
\altaffiliation[Currently at ]{National Institute of Standards and
  Technology, Boulder, Colorado 80305}
\email{janus@boulder.nist.gov}
\author{Klaus M{\o}lmer}
\affiliation{Danish National Research Foundation Center for Quantum Optics, Department of Physics and
Astronomy, University of Aarhus, DK-8000 Aarhus C., Denmark}
\author{Lars Rippe}
\author{Stefan Kr{\"o}ll}
\affiliation{Department of Physics, Lund Institute of Technology, Box 118, S-221 00 Lund, Sweden}

\begin{abstract}
  Due to inhomogeneous broadening, the absorption lines of
  rare-earth-ion dopands in crystals are many order of magnitudes
  wider than the homogeneous linewidths. Several ways have been
  proposed to use ions with different inhomogeneous shifts as qubit
  registers, and to perform gate operations between such registers by
  means of the static dipole coupling between the ions.
  
  In this paper we show that in order to implement high-fidelity
  quantum gate operations by means of the static dipole interaction,
  we require the participating ions to be strongly coupled, and that
  the density of such strongly coupled registers in general scales
  poorly with register size. Although this is critical to previous
  proposals which rely on a high density of functional registers, we
  describe architectures and preparation strategies that will allow
  scalable quantum computers based on rare-earth-ion doped crystals.
\end{abstract}


\maketitle

\section*{\label{sec:introduction} Introduction }
Several proposals have been made for quantum computers based on  rare-earth ions
embedded in cryogenic crystals
\cite{ohlssona02:quant_comput_hardw_based_rare,lukin00:quant_entan_optic_contr_atom,pryde00:solid_state_coher_trans_measur}.
In the REQC (Rare-Earth Quantum Computing) proposals, qubits are identified with frequency channels
appropriately selected within the inhomogeneous absorption profile of the dopant ions in the crystal.
Full $N$-bit quantum registers occur at random in the material where a collection of ions with the
selected absorption frequencies happen to be physically close enough to be able to control each other
by the interaction between their excited state dipoles. The probability of finding a number $N$ of
such interacting particles has a poor $\nbar^N$ scaling (exponential exp(-$|$log($\nbar$)$|N$)), where
$\nbar$ is the occupancy of the selected frequency range within the spatial interaction volume of the
other ions.

In this paper we discuss means to achieve scalability in the rare-earth-doped crystals. In Section I,
we briefly review the REQC idea for one and two-bit gates in the rare-earth-ion quantum computer. In
Section II, we discuss quantitatively the scaling problem, and we suggest to use a bus-architecture
where a single ion is dedicated to provide the communication within the register and to ensure that
there is with large probability precisely one ion present at each frequency channel. In Section III,
we address the possibility of working with only a single instance rather than a large ensemble of
identical quantum registers. For this purpose it is necessary to have a single species with very high
read-out efficiency (single-ion detection). Section IV concludes the paper.

\section{One- and two-qubit gates}
\label{sec:rare-earth-quantum}

\begin{figure}[htbp]
  \centering
  \subfigure[]{%
    \includegraphics[width=0.3\linewidth]{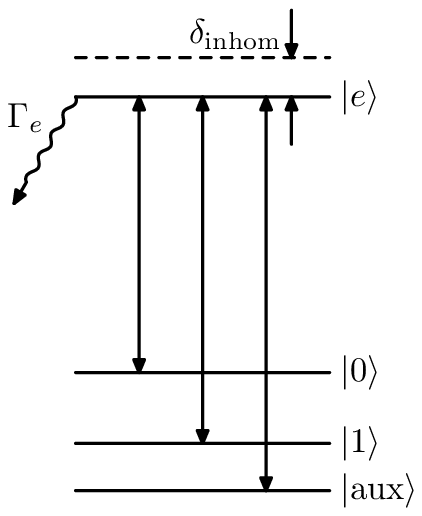}%
    \label{fig:levelscheme}
  }
  \subfigure[]{%
    \includegraphics[width=0.65\linewidth]{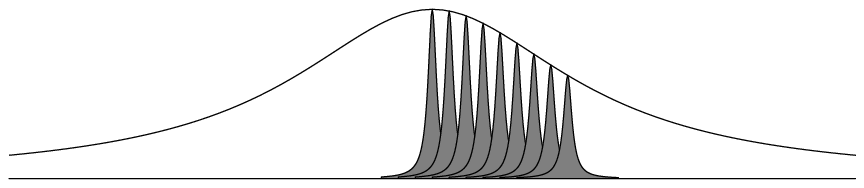}%
    \label{fig:hominhom}
  }
  \caption{Simplified level scheme for rare-earth-ion
    \subref{fig:levelscheme}. Inhomogeneous broadening
    \subref{fig:hominhom}.
  }
\end{figure}
In this section we briefly introduce the rare-earth quantum computing proposal, described in more
detail in Refs.
\cite{nilsson02:initial_exper_concer_quant_infor,ohlssona02:quant_comput_hardw_based_rare,roos03:robus_quant_comput_with_compos,wesenberg03:robus_quant_gates_a_archit,rippenilsson}.

\begin{figure}[htbp]
  \centering
  \includegraphics[width=\linewidth]{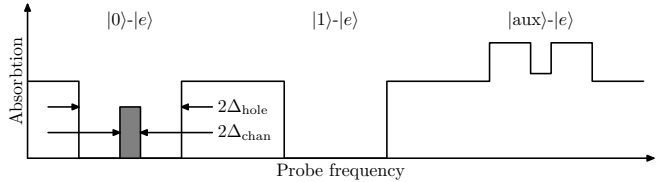}
  \caption{Schematic structure of channel absorption profile}
  \label{fig:channel}
\end{figure}

For the purpose of this article, we consider rare-earth-ions with three metastable ground state
hyperfine levels, labeled $\ko$, $\kl$ and $\kaux$, which are all coupled by optical transitions to an
excited state $\ke$ (Fig. \ref{fig:levelscheme}). The excited state $\ke$ is inhomogeneously shifted
by up to several GHz. Since the homogeneous line width of the optical transitions can be as low as
kHz, this allows us to address a very large number of \emph{channels}: frequency selected subgroups of
the ions identified by their inhomogeneous shift (Fig. \ref{fig:hominhom}).
Different channels are selected at sufficiently different frequencies to allow unambiguous addressing,
and each channel is separated from the broad inhomogeneous profile by spectrally hole burning an
interval around the frequency channel, so that only relatively narrow structures in frequency space
are addressed by laser fields (Fig. \ref{fig:channel}). It is now possible to use the optical
frequency to selectively perform quantum gates on the individual qubits, i.e., drive transitions
between the qubit-levels $\ko$ and $\kl$ via the excited state $\ke$. To carry out two-bit gates in
the rare-earth quantum computer we use that the ions have a controlled interaction due to the static
dipole coupling between the $\ke$-states. The strength of this coupling is determined by the
dipole-moments and it differs for different ion pairs as it depends on the relative position of the
ions. By using the dipole blockade effect \cite{jaksch}, however, it is possible to implement reliable
quantum gate operations that act independently of the precise value of the strength of the coupling
between the involved ions, provided that this strength exceeds a certain threshold on the order of the
intervals vacated by hole-burning around the channels. We will consequently refer to pairs of ions as
being coupled or not depending on whether their coupling strength exceeds the threshold value. A
two-bit quantum gate is now performed by first driving the ions on one qubit frequency channel
resonantly from one of the qubit levels into the excited state. The application of radiation resonant
with another qubit channel frequency will now excite the ions resonantly, provided the ions in the
first qubit are not excited, whereas the excitation is shifted out of resonance if the first ion is
excited. This results in a coherent evolution of the ions in the second qubit conditioned on the
initial qubit quantum state of the first ion as proposed in
\cite{nilsson02:initial_exper_concer_quant_infor,ohlssona02:quant_comput_hardw_based_rare}. The ions
are only addressed according to their optical excitation frequencies, and there will in general be a
very large number of ions having the same frequencies. This is a desired feature of the proposal,
since the read-out, as in the case of the NMR quantum computer is making use of the macroscopic signal
from the entire ensemble. Unlike the NMR computer, however, the initial state is fully controlled, and
all the ensemble members ideally perform the same unitary evolution. The read-out is thus obtained
from a pure state ensemble. In
refs.\cite{roos03:robus_quant_comput_with_compos,wesenberg03:robus_quant_gates_a_archit}, it was
further proposed to use composite pulses and suitably tailored continuous pulses to make the gates
robust against the small frequency dispersion of the optical frequencies within the selected channels.

In weakly doped crystals like those used in recent experiments on Pr, Eu and Tm doped crystals
\cite{holeburning,rippenilsson,longdell,longsellar,fraval,seze} the difference in dipole moment is
sufficiently large that each qubit ion has several neighboring ions that it can control. Each of these
neighbors can in turn control several other ions etc. Consequently, from every ion there are branched
chains of ions, where each ion in the chain can control its nearest neighbors. Thus, there is a large
number of potential quantum processors in any crystal. Such a processor, however, operates with
specific qubit frequencies located at random positions within the inhomogeneous absorption profile,
and only if the experiment has the ability to detect single rare-earth-ions, it will be possible to
operate. Hence a good deal of theoretical and experimental work has focussed on multiple instance
implementations
\cite{wesenberg03:robus_quant_gates_a_archit,nilsson02:initial_exper_concer_quant_infor,nilssonlevin,roos03:robus_quant_comput_with_compos,holeburning,rippenilsson,longdell,longsellar,fraval,seze,janusw}.
The multiple instance implementations are characterized by a choice of definite values for the qubit
frequency channels followed by an optical pumping/holeburning to identify the instances where these
selected frequency channels all are occupied by (at least) one ion.
\section{Ensemble quantum computing with bus-ions}

\subsection{Use of a bus ion for communication within a quantum register}

For a two-bit gate to be possible the ions must be interacting, i.e., they have to be within a
critical distance $R$ of each other. In the original proposal for computing with doped ions, any pair
of ions should be within this distance, i.e., they should be within a ball of diameter $R$ Without
changing the physical implementation of the quantum computer, it is possible to increase the number of
functioning registers, by assigning the role to one of the ions to take care of all communication in
the register so that two-bit gate operations between any pair of ions takes place using only their
mutual interactions with the dedicated bus-ion \cite{wesenberg03:robus_quant_gates_a_archit}.  Using
the intervening bus-ion, all ions should only be within a distance of $R$ from the bus, i.e., the ions
should be within a sphere of diameter $2R$, centered around the bus. This enlargement of the
interaction volume by a factor 8 thus enlarges by a factor of approximately $8^{n-2}$ the number of
n-bit registers in the crystal ($n=2$ ions have to be within distance $R$ of each other, and only for
three ions and more can the ions have larger mutual distances).

\subsection{Use of a bus ion for hole burning and initialization of a quantum register}

A quantum register is only operational if all qubit channels are populated by a single ion, and in the
original proposal, hole burning is thus applied both to define the frequency channels and to remove
ions which do not have sufficient interactions with the other qubits, either because they are not
physically close enough or simply because there is no ion at that frequency in the vicinity at all. In
the bus-architecture just outlined, we must thus secure that every ion interacts with the bus,
otherwise the whole register will be erroneous, and all ions should be removed by the hole burning
procedure: We know that the excited bus shifts the channel frequencies, and thus protects ions from
hole burning, and therefore hole burning at each channel frequency in the presence of an excited bus
will leave all good registers intact and only vacate the useless qubit channels. It is also possible
to remove the bus itself if it is not surrounded by the appropriate set of register ions. A specific
procedure to obtain this configuration is proposed in Ref.
\cite{ohlssona02:quant_comput_hardw_based_rare} and partly experimentally demonstrated in
\cite{rippenilsson}. This procedure relies on repeatedly pumping unusable ions out of their qubit
states, thus transferring them with certainty to their $\kaux$-state. After these hole burning
operations we are left with a stable crystal which contains a number of disjunct, independent and
perfect quantum computer instances.

\subsection{Scaling of the proposal}

Without single ion read-out capabilities, the ensemble quantum computing proposal relies on a large
number of equivalent quantum processors in a single crystal, so that the individual qubits can be read
out by the macroscopic absorption or dispersion properties of the crystal around the qubit
frequencies. Let us assume that the density of channel members and requirements on coupling strength
is so that the frequency interval assigned to each channel is on average populated by $\nbar$ ions. At
the low doping concentrations, considered in the initial proposal
\cite{ohlssona02:quant_comput_hardw_based_rare}, $\nbar$ will be much smaller than unity, but we shall
not restrict ourselves to that regime here. For a random doping, it is reasonable to assume the
inhomogeneous shifts of the ions to be spatially uncorrelated, corresponding to a Poissonian
distribution of the number of ions in a qubit-channel. But, only if there is precisely a single ion in
each channel, we will have a useful register. Hence, let us consider the probability $P_N$ that a
given ion is coupled to precisely one ion in each of the $N$ channels constituting our register. The
probability of having $m$ ions in a channel is $\exp(-\nbar) \nbar^m/m!$, and for the relevant value
$m=1$, we get
\begin{equation}
  \label{eq:pndef}
  P_N=(e^{-\nbar}\ \nbar)^N \le e^{-N},
\end{equation}
where the maximum is obtained for $\nbar = 1$. Although high doping gives a better scaling than a
really low concentration, the proposal still has a poor, exponential scaling with respect to $N$ of
the probability and hence the total number of instances in the crystal.
We note, however, that if a further initialization step can be made so that channels with any non-zero
occupancy can be manipulated to contain exactly a single ion, each channel can be fixed with a
probability which is unity minus the zero-occupancy population, and starting again with the Poisson
distribution with a mean occupancy of $\nbar$, the N-bit register will be operational with a
probability
\begin{equation}
  \label{eq:pndefany}
  P'_N=\left(1-e^{-\nbar}\right)^N,
\end{equation}
so that $P'_N\approx1$ if we can ensure  $\nbar\approx\log(N/(1-P'_N))$.

\subsection{Controlling the number of ions in each frequency channel by optical interactions}

\begin{figure}[htbp]
  \centering
  \includegraphics{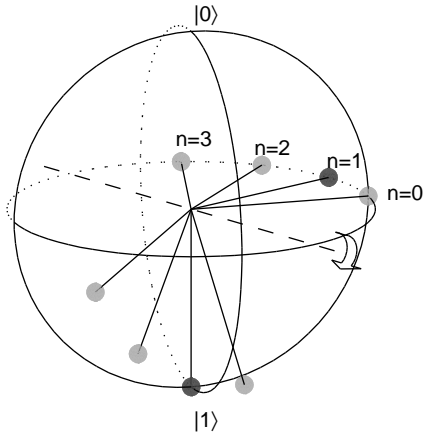}
  \caption{
    Graphical illustration of the evolution of the central qubit
    during the process described for removing superfluous ions.
  }
  \label{fig:blochfig}
\end{figure}

We now present a method to reduce the number of ions in each frequency channel to unity using only
global operations on the entire crystal, i.e., on an ensemble of systems which initially has different
occupancies of the frequency channels, and which should eventually consist of a large number of
identical registers. This can be done by inducing a sequential interaction between the bus-ion and the
ions in each of the frequency channels according to the following protocol:

The central ion is prepared in the superposition state $(\ko+\kl)/\sqrt{2}$, and a phase shift of
$\exp{i \alpha}$ is applied to all ions in the channel conditioned on the central qubit being in the
$\kl$-state. The $\ko$-state amplitude is transferred to the excited state of the central ion, and
subsequently, a resonant transfer is driven from the $\ko$-state of the other ions into their excited
state and back again with a phase difference of the two fields of $\alpha$. Since the resonant process
only takes place if the central ion does not shift the transition frequency, i.e., if it is in the
$\kl$-state, the result of the transfers is to leave the channel ions unchanged in the $\ko$-state and
the central ion in the state $(\ko+\exp(i n \alpha)\kl)/\sqrt{2}$ where $n$ is the number of channel
members coupled to the ion.
We aim at protecting the $n=1$ state, and a rotation of the central ion by $\pi/2$ around an axis with
azimuthal angle $\alpha+\pi/2$ on the Bloch sphere will map the combined state of the central and
channel ions to
\begin{equation}
  \label{eq:measstate}
  \ket{\psi}=\left(
  \sin\left(\frac{(n-1)\alpha}{2}\right)\ko
  +\cos\left(\frac{(n-1)\alpha}{2}\right)\kl
  \right)\ \ko^{\otimes n},
\end{equation}
up to a global phase, as illustrated in Fig. \ref{fig:blochfig}.

A \textsc{cnot} on the channel ions conditioned  on the state of the
central qubit will transfer the channel ions to a state
$\ket{\psi}^{\otimes n}$ which has non-vanishing overlap with
$\ko^{\otimes n}$ only if $n \neq 1$.
We now transfer the channel ions in their $\ko$-state to an excited state (preferably with high
branching ratio to $\kaux$) with a probability amplitude of $\beta$ and wait for the ions to decay.
After this, we perform another \textsc{cnot} to return all channel
ions to their $\ko$ state and re-initialize the bus, after which we
repeat the process until all superfluous channel representatives have
been removed.

The danger of the above process is that all ions in an instance can
decay to their $\kaux$-state simultaneously. The probability of this
happening is $\beta^n$, compared to the probability $n\, \beta^{n-1}
(1-\beta)$ of $n-1$ ions decaying, so that by choosing $\beta$
sufficiently small, we can achieve any desired efficiency at the price
of increasing the number of necessary repetitions.

By applying the above procedure repetitively, we can remove superfluous ions and monotonically
increase the probability of having precisely one atom per channel.

In order to obtain a scalable implementation, $\nbar$ must be larger than unity, but it only grows
logarithmically with the size of the register. To obtain $P_N=90\%$ working instances of an
$N=10(100)$ qubit system, \eqref{eq:pndefany} implies that $\nbar$ must be larger than $4.6(6.9)$.
Such comparatively high values are for example obtained in stoichiometric crystals and in some of
these the rare earth transitions can still have remarkably narrow line widths and long coherence times
in spite of the high concentration of active ions \cite{shelbymac}. The complexity and the number of
physical operations during the initialization process of course grows with the required register size.

The value of $\nbar$ in a given system depends on two parameters: the ion concentration and the
coupling strength required to perform gate operations. In our above analysis, we have focussed on the
case of a very large coupling strength (some MHz) between ions, requiring the ions to be close and
hence limiting the number of useful instances. It should be noted that a different mode of operation
has been proposed \cite{smallcoup}, in which much smaller interaction strengths (some kHz) are
required, and where refocussing techniques are applied to ensure interaction between ions with only
the appropriate interaction strengths. As pointed out in \cite{smallcoup}, such low interaction
strengths make ions from a much larger volume available for quantum gates, but it also implies slower
gates, and hence the lifetime of the excited state quickly becomes a serious issue. For further
reference, we prove in the appendix, that in order to maximally entangle a pair of ions with the
excited state dipole-dipole interaction, the total population, $\pextot$, in the excited state during
the gate operation must obey $\int \pextot dt\ >\ 2/g$, where $g$ is the dipole coupling strength.
I.e., in order to obtain a high fidelity gate operation, the interaction strength must be orders of
magnitude larger than the excited state decay rate.

\section{Single instance quantum computing with a "designated read-out ion"}

Although the above analysis leaves grounds for some optimism concerning the implementation of quantum
computing with not too large registers in the ensemble mode, we believe that full scalability will be
more realistically obtained in a single instance implementation. Single ion detection has been
demonstrated for Pr and Eu doped crystals \cite{bartko}, but high-fidelity qubit state-selective
read-out remains difficult because of the $>$ 100 $\mu$s upper state lifetimes in combination with the
absence of a state which can be cycled repetitively as the shelving technique applied in trapped ion
studies \cite{shelving}. To solve this problem we propose to add an additional physical system to the
crystal designated only for the readout process. In similarity with the bus ion approach, where the
bus ion needs to have strong ion-ion interaction while, e.g., long coherence times is less important,
the physical requirements for a read-out ion will be different from the qubit ion requirements. For
example, as long as the read-out ion is able to provide information about the state of the qubit ions,
there are no requirements on its coherence times and this opens for a large number of possible
read-out system. The readout system could be a molecule, a colour centre, an ion or some other kind of
suitable physical system, but for simplicity we will henceforth refer to the read-out system as the
"read-out ion".

\subsection{Read-out ion properties}

We will now describe a single instance REQC implementation based on having a designated read-out ion.
The general idea is to have the read-out ion (system) so close to a qubit ion that it can be shifted
in and out of resonance with a laser field by transferring the qubit ion between the ground and
excited state. In this way the read-out system fluorescence signal provides information about the
state of the qubit ion.

A reasonable set of requirements on a read-out ion (system) could be the following:

1. The read-out system must not be trapped in a non-fluorescent state.

2. It should produce a number of output photons/second sufficient for detecting a single readout
system.

3. It should be possible to have a qubit ion close enough to the readout system to shift it off
resonance by an interaction that can be turned on and off.

4. The presence of the readout system should not significantly decrease the qubit coherence time.

5. Readout ion operation should not induce state changes of neighboring qubit ions.

Assume that promoting the qubit ion to the excited state will shift the read-out ion into resonance
with a laser field at frequency  $\nu_0$. To make the discussion more transparent it will be assumed
that the qubit ions are Eu ions doped into an Y$_2$SiO$_5$ crystal. The excited Eu state that appears
best suited for qubit operations has a life time of 2 ms. Requiring at least 100 detected photons and
assuming a 1 $\%$ photon detection efficiency, the read-out ion should emit at least 1 photon/200 ns.
If we further require that the read-out process should work in 99 $\%$ of all cases it must be
possible to cycle the readout ion of the order of $10^6$ times before it is trapped in a
non-fluorescent state. Regarding condition number 2 above, the lifetime limited line width with a 200
ns excited state is about one MHz. We assume that the readout system has a dipole moment difference
$\Delta \mu$, so that one nanometer from a Eu ion, the electric dipole-dipole interaction would induce
a frequency shift of $\Delta\mu/(10^{-30}Cm)$ GHz (or ~3.5 GHz/Debye). The dipole moment difference
for Eu in Y$_2$SiO$_5$ is $0.8 10^{-31}Cm$ (0.023 Debye). If the readout system has a similarly small
dipole moment difference of $0.8 10^{-31}Cm$, the line shift one nm away from an Eu qubit ion would be
close to 100 MHz.

\subsection{Qubit read-out}

The system could typically operate with one laser interacting with the qubits (\emph{qubit laser}) and
one laser interacting with the readout system (\emph{readout laser}). To read out a qubit (\emph{qubit
1}) close to the readout ion, the readout laser is tuned to frequency $\nu_0$ and a \emph{qubit laser}
$\pi$-pulse is applied to the \emph{qubit 1} $|0\rangle-|e\rangle$ transition. If the readout ion
fluoresces, \emph{qubit 1} was in the $|0\rangle$ state.

To read out a neighbor (\emph{qubit 2}) of \emph{qubit 1}, which is not directly coupled to the read
out ion, we may either use a quantum gate to swap qubit states 1 and 2, or in a sequential read-out of
all ions, \emph{qubit 1} can be pumped into the $|0\rangle$ state and a CNOT operation with
\emph{qubit 2} as control qubit is carried out to transfer the \emph{qubit 2} state to \emph{qubit 1}.
\emph{Qubit 1} can now be read out as described above. It is straightforward to extend the approach to
consecutively read out of all qubits in the register.

\subsection{Initiation of single instance computer}

We will briefly discuss a design and initiation procedure for the single ion readout scheme. To only
read out a single instance it is necessary to have sufficiently low readout ion concentration that
there is only one fluorescing readout ion interacting with the laser radiation at frequency $\nu_0$
within the volume observed. Further, before operating the system it has to be fully characterized such
that the readout ion frequency and all qubit transition frequencies are known. This is done by first
scanning the readout laser across the readout ion inhomogeneous line width until readout ion
fluorescence is detected. Then the readout laser remains tuned to the readout ion resonance. Now the
frequency of the qubit laser is tuned across the inhomogeneous qubit transition and $\pi$-pulses are
applied to the crystal. Once the fluorescence stops, the transition frequency of a qubit (\emph{qubit
1}) that controls the readout ion has been found. By observing the readout ion fluorescence (and
applying appropriate operations) the energy level structure of \emph{qubit 1} can now be mapped out.
By promoting \emph{qubit 1} to the excited state, the readout ion transition frequency $\nu_0$ when
\emph{qubit 1} is excited, can be found. Finding the next qubit, \emph{qubit 2}, is done in a way very
similar to finding \emph{qubit 1}. First \emph{qubit 1} is put in the $|0\rangle$ state. The readout
laser is tuned to frequency  $\nu_0$. Now the frequency of the qubit laser is again tuned across the
inhomogeneous qubit transition and $\pi$-pulses are applied to the crystal. After each $\pi$-pulse an
additional $\pi$-pulse is applied to the \emph{qubit 1} $|0\rangle -|e\rangle$ transition. This will
start the fluorescence signal unless the \emph{qubit 1} transition frequency has been shifted by
\emph{qubit 2}. Once the fluorescence does not start after the \emph{qubit 1} $|0\rangle -|e\rangle$
$\pi$-pulse has been applied, the \emph{qubit 2} frequency is found. The scheme can readily be
extended further along the chain using an analogous procedure. The initiation procedure described
above requires high efficiency $\pi$-pulses. However, highly efficient $\pi$-pulses for these systems
have indeed been both designed \cite{roos03:robus_quant_comput_with_compos,janusw} and experimentally
demonstrated \cite{holeburning,rippenilsson,seze}.

\subsection{Physical candidate for readout ions}

An actual physical readout system needs to be identified, and the 4f-5d transition in Ce$^{3+}$ has
been proposed as a potentially viable candidate by Guillot-No\"{e}l \cite{noel}. For Ce$^{3+}$ in
YPO$_4$ the homogeneous line width of the 4f-5d transition is about 50 MHz and the 5d state lifetime
is about 20 ns \cite{holeburning}. The corresponding values for the host Y$_2$SiO$_5$ could be
expected to be similar. Ce$^{3+}$ has a zero nuclear magnetic moment and therefore the ion
fluorescence can not be quenched by decay to a hyperfine state not interacting with the laser at
frequency $\nu_0$. The dipole moment difference is unknown but can be expected to be 0.1 Debye or
larger \cite{meltzer,kaplyanskii}. Thus based on these actual and anticipated data, the Ce$^{3+}$ ion
appears to be an interesting readout system candidate.

Single-ion readout does not make the system fully scalable in itself. The system increases in size by
adding new members to the (possibly branched) chain of qubit ions. Although there are many frequency
channels available within the inhomogeneously broadened qubit transition line, a new qubit will
eventually happen to have the same frequency as one of the ions already in the chain and these two
ions then cannot be individually addressed. The problem of coinciding transition frequencies can be
removed by mounting (closely spaced) electrodes on the crystal. Eu in Y$_2$SiO$_5$ has a Stark
coefficient of 35 kHz/(V/cm) \cite{graf}. Thus a 1 MV/cm field would shift the transition frequency 35
GHz, a detuning much larger than the inhomogeneous transition line width. The part of the qubit chain
on which the operations should be carried out can be selected by applying a voltage on the appropriate
electrode. All parts of the chain not close to this electrode remain unaffected by light pulses. For
such a design the system appears to be fully scalable. A more "algorithmic" solution to the problem of
degeneracies among the qubit frequencies, is to replace all single qubit operations by two-bit
operations, so that operations are applied to "the qubit with frequency $\nu_i$, sitting next to a
qubit with frequency $\nu_j$".

\section{Conclusion}

Experimental work to develop rudimentary precursors of quantum computer hardware in rare earth ion
doped crystals have been based on using many instance ensembles of quantum computers in weakly doped
crystals. These approaches are, however, not readily scalable to arbitrary large number of qubits. In
this manuscript scalability requirements for rare-earth-crystal-based quantum computer hardware have
been analysed and in particular two scalable approaches for rare earth quantum computing have been
introduced. One of these is an ensemble-based many instance approach where highly doped crystals (for
example stoichiometric crystals \cite{shelbymac}) are used to obtain sufficiently large number of
closely situated ions that are able to control each other. The other approach is a single instance
scheme where specially designed read-out ions are used to read out the values of the qubit ions.

This work has been supported by the ESQUIRE project within the IST-FET program of the EU, the Swedish
Research Council and the Danish National Research Foundation.

\section{Appendix}

The dipole gate makes use of the interaction between two ions in their excited states to which they
may be selectively promoted from one of the qubit levels. When both ions are excited, the interaction
causes an energy shift, which accumulates a phase factor, leading to a non-trivial two-bit gate
operation. As suggested in \cite{ohlssona02:quant_comput_hardw_based_rare,jaksch}, it is enough to
excite one of the ions and use the shift in resonance frequency to actually avoid excitation of the
other one. This implementation is robust against variations of the coupling strength. Since the
excited state has a finite lifetime, it is worth looking for proposals that minimize the total time
spent by atomic population in the excited states, and one may for example speculate if there is a way
to off-resonantly couple both ions, so that the excited state interaction is effective without any of
the ions significantly populating that state. We shall now prove that this is unfortunately not
possible.

We start by considering the general setting of two quantum
subsystems described by Hilbert spaces $\hilb_a$ and $\hilb_b$.
If the combined system is described by a pure state $\ket{\Psi}$ in
the combined Hilbert space $\hilb=\hilb_a \otimes \hilb_b$, the degree
of entanglement between the two systems is quantified by the \emph{von
Neumann entropy} of the reduced density matrix of either one of the
systems: $E=S(\rho_a)=S(\rho_b)$.

Calculating $E$ is most easily done if we Schmidt-decompose
$\ket{\Psi}$ as
\begin{equation}
  \label{eq:phsischmidt}
  \ket{\Psi}=\sum_i c_i \ket{v_i} \ket{u_i},
\end{equation}
where the $c_i$ are real and non-negative, and $\{ \ket{v_i} \}$ and
$\{ \ket{u_i} \}$ are orthonormal bases of $\hilb_a$ and $\hilb_b$,
respectively.
In terms of the Schmidt decomposition, we have $\rho_a =\sum_i c_i^2
\ket{v_i}\bra{v_i}$, so that
$E=S(\rho_a)=- \sum_i c_i^2 \log_2(c_i^2)$.

To maintain the Schmidt decomposition \eqref{eq:phsischmidt}
during evolution, we must allow all of $c_i$, $\ket{u_i}$, and
$\ket{v_i}$ to be time-dependent.
The time derivative of the Schmidt coefficients is found to be
\begin{equation}
  \label{eq:cj2deriv}
  \left. \dd{t} c_i^2 \right|_{t=0}
  =\left.
    \dd{t} \braket{v_i(0) | \rho_a(t) | v_i(0)}
  \right|_{t=0},
\end{equation}
since $\rho_a=\sum_i c_i^2 \ket{v_i}\bra{v_i}$ and $\{\ket{v_i(t)}\}$
is an orthonormal basis at all times.
To evaluate the derivative, we perform the partial trace over $\hilb_b$
in the $\ket{u_i(0)}$ basis:
\begin{equation}
  \label{eq:dtcj2}
  \left. \dd{t} c_j^2 \right|_{t=0}
  =2 c_j(0) \im \braket{v_j(0) u_j(0) | \ham | \Psi(0)},
\end{equation}
using the Schr\"odinger equation and that $\braket{v_j u_i | \Psi} =
c_j \delta_{i,j}$.
Finally, we compute the time derivative of $E$ according to
\eqref{eq:dendtsym}, and using that $\sum c_i^2=1$ we find:
\begin{equation}
  \label{eq:dendtsym}
  \dot{E}=2 \sum_{i, j} \log_2\left(\frac{c_j}{c_i}\right) c_i c_j
  \im( \braket{u_i v_i | \ham | u_j v_j}),
\end{equation}
where we note that the diagonal terms do not contribute.

We see that this expression for $\dot{E}$ is additive in $\ham$: $\dot{E}(\ham_a
+\ham_b)=\dot{E}(\ham_a)+\dot{E}(\ham_b)$, and that
 terms of the form $\id \otimes \ham_a$ corresponding to local
 operations do not create entanglement:
$\dot{E}(\id \otimes  \ham_a) =0$
as we would expect.

According to this argument, the only part of the two-qubit Hamiltonian
in REQC capable of producing entanglement is the coupling term $\ham_c=g
\ket{ee}\bra{ee}$.
The instantaneous rate of creation of entanglement only depends on the
instantaneous state of the ions, which, in turn, is controlled by the
applied fields.

Using that $a+b\ge2\sqrt{ab}$ for any non-negative $a$ and $b$, we see that
\begin{equation}
  \label{eq:wineq}
  w_i \equiv\tfrac{1}{2}   \left(
    |\braket{u_i|e}|^2+|\braket{v_i|e}|^2
  \right)  \ge
  | \braket{u_i v_i|ee}|,
\end{equation}
where we have introduced $\{w_i\}$, which are seen to fulfill $\sum_i
w_i=1$ as  $\{ \ket{u_i} \}$ and $\{ \ket{v_i} \}$ are orthonormal
bases.
Inserting $\ham_c$ into the expression \eqref{eq:dendtsym}, the
inequality \eqref{eq:wineq}
allows us to establish the following bound
on the rate of entanglement creation:
\begin{equation*}
  \frac{|\dot{E}|}{g} \le
  2 \sum_i  c_i^2 w_i \sum_j f(\theta_{i,j})  w_j,
\end{equation*}
where we have parameterized $(c_i,c_j)$ as $(\sin(\theta_{i,j}),
\cos(\theta_{i,j})) (c_i^2+c_j^2)$, and introduced
$f(\theta)=|\log(\tan(\theta))| \sin(\theta) \cos(\theta)$.
A numerical analysis shows that $f$ is bounded by $0.478\ldots$, and hence
\begin{equation}
  \label{eq:dentdtfinal}
  |\dot{E}| <  \frac{1}{2} \ \pextot \ g,
\end{equation}
where
$\pextot
= \braket{\Psi|\left(\id \otimes \ke\be + \ke\be \otimes \id \right)|\Psi}
= 2 \sum_i w_i c_i^2$ is the total excited state population.

The implications of Eq.~\eqref{eq:dentdtfinal} are clear: in order to be able to obtain one unit of
entanglement, as required for a universal two-qubit gate operation, the system must necessarily suffer
an integrated excitation $\int \pextot dt\ >\ 2/g$, where $g$ is the dipole coupling strength.


\end{document}